# Photonics design tool for advanced CMOS nodes

Luca Alloatti[1] ✉, Mark Wade[1,2], Vladimir Stojanovic[3], Milos Popovic[2], Rajeev Jagga Ram[1]
[1]Massachusetts Institute of Technology, Cambridge, MA 02139, USA
[2]Department of Electrical, Computer and Energy Engineering, University of Colorado, Boulder, CO 80309, USA
[3]Department of Electrical Engineering and Computer Science, University of California, Berkeley, CA 94720, USA
✉ E-mail: alloatti@mit.edu

**Abstract:** Recently, the authors have demonstrated large-scale integrated systems with several million transistors and hundreds of photonic elements. Yielding such large-scale integrated systems requires a design-for-manufacture rigour that is embodied in the 10 000 to 50 000 design rules that these designs must comply within advanced complementary metal-oxide semiconductor manufacturing. Here, the authors present a photonic design automation tool which allows automatic generation of layouts without design-rule violations. This tool is written in SKILL, the native language of the mainstream electric design automation software, Cadence®. This allows seamless integration of photonic and electronic design in a single environment. The tool leverages intuitive photonic layer definitions, allowing the designer to focus on the physical properties rather than on technology-dependent details. For the first time the authors present an algorithm for removal of design-rule violations from photonic layouts based on Manhattan discretisation, Boolean and sizing operations. This algorithm is not limited to the implementation in SKILL, and can in principle be implemented in any scripting language. Connectivity is achieved with software-defined waveguide ports and low-level procedures that enable auto-routing of waveguide connections.

## 1 Introduction

Large-scale integrated systems require multiple process steps and mask layers to define etch and deposition patterns, implants and metallisation. In electronic complementary metal-oxide semiconductor (CMOS) manufacturing, well-constructed design rules (DRs) ensure that billions to trillions of nanoscale components can be fabricated simultaneously with high yield and performance. Similarly, photonic integrated circuits (PICs) with hundreds or thousands of individual components, as well as monolithically integrated photonic circuits [1–4] now rely on the same paradigm. However, while modern electric design automation (EDA) tools can automatically generate electronic circuit layouts without design-rule check (DRC) violations starting from abstract hardware descriptions, there is no equivalent tool infrastructure for PICs.

Here, we present a photonic design automation (PDA) tool that allows designers to define optical structures using abstract and technology-independent layers which are then automatically mapped onto DRC-clean mask design levels. We illustrate the PDA tool with photonics implemented in a 45 nm CMOS microelectronics silicon-on-insulator (SOI) process.

Within the 45 nm CMOS node we have recently demonstrated optical transceivers next to million-transistor electrical circuits without requiring any changes to the fabrication process flow (a paradigm we termed 'zero-change CMOS' photonics) [4, 5]. The on-chip transmitter consists of a pseudorandom bit sequence (PRBS) generator, a spoked-ring carrier-depletion modulator [6] and electrical driver with serialiser while a receiver consists of a silicon-germanium photodetector, a transimpedance amplifier and a digital back-end with a sampling-scope and bit-error-ratio (BER) tester functionality. The transmit/receive blocks operate with an electrical power consumption of <0.3 pJ/bit at a data rate of 3.5 Gb/s and 2.5 Gbit/s [5], respectively (Fig. 1).

We first present an overview of the available PDA tools. We then discuss the user interface (UI) and the use of abstract design levels. Finally we discuss the manufacturing DRs, their implications for photonics and demonstrate the PDA approach to removing DRC violations.

## 2 Photonic design tools, state-of-the-art

So far, a dominant part of integrated photonics research has exploited silicon or InP foundries running customised fabrication processes for photonics which have offered photonic nanofabrication services and multi project wafer runs [7–12]. The majority of these foundries however have offered a simplified CMOS flow without transistors, typically one to four masks and a minimal amount of DRs. As a consequence, the majority of the photonic design tools developed to date focus on integrating electromagnetic simulation tools, waveguide routing, embedded behavioural models for system simulation and scripted or graphical layout generation [13–24]. The automated part of the design is typically limited to place-and-route of photonic components, while the layout is drawn directly without DRC violations. Concerning electronic-photonic design, Mentor Graphics [25] recently announced a collaboration with the photonic design and simulation companies PhoeniX Software and Lumerical Solutions [26]. Finally, Cadence® has been utilised for designing photonic components by Luxtera, and has been utilised in our group since 2006 [27, 28], however post-layout layer generation, automatic Manhattan discretisation, automatic DRC cleaning and photonic auto-routing had not been implemented.

## 3 Cadence-based photonic design tool

We have developed a PDA tool based on a mainstream front-end electronic design software, Cadence® Virtuoso and written in the native Cadence® scripting language, SKILL. Layouts are generated completely through scripts. SKILL code is used to define parameterised cells (known as pCells) with a hierarchical and modular structure, as well as procedures (or functions) which

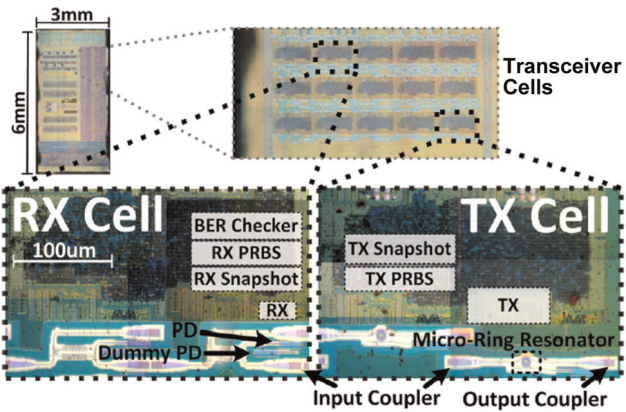

**Fig. 1** *Optical transceiver monolithically integrated with electrical circuits in 45 nm 12SOI IBM technology [5]*

Full 3 mm×6 mm die comprises individual transmit/receive cells, each containing a PRBS generator, modulator driver, modulator, SiGe detector and a BER tester

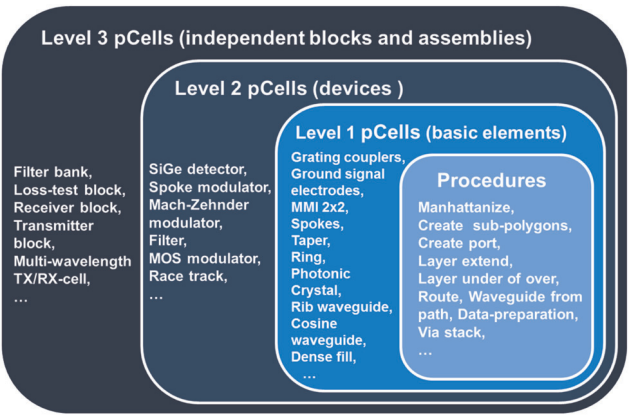

**Fig. 2** *Hierarchical structure of the code*

At the lowest level there are procedures which extend the functionality of Cadence. Parametrised cells (pCells) are built hierarchically starting from basic elements, to single devices, to chip assemblies execute particular tasks, such as generating a waveguide starting from a path, aligning waveguide ports together or removing design-rule violations. SKILL is a sophisticated language which very well satisfies the needs of CMOS electronics designers and is an ideal candidate for an expansion to photonic design. In fact, built-in functions allow manipulating layouts, shapes, instances and connectivity properties, just to list a few. However, the SKILL environment comes with two major limitations, namely the impossibility of defining polygons with more than 4000 points (a legacy limitation), and the absence of built-in waveguide-ports functionalities. The former limitation is not an issue when designing transistors: these are typically made of rectangles, and can be defined with a very small set of points. Photonic components instead, such as rings or bends, have dimensions of a few micrometres at least, leading to polygons violating this limit. Moreover the concept of waveguide-port has no analog in the electronic world: while the connectivity of two wires requires just a metallic link between them of (almost) any shape and direction, in the context of waveguides one must take account of port width, location and orientation. Since a typical layout consists of several instances of individual master cells, the waveguide ports should automatically align and reorient according to the properties of each instance. Both of these limitations have been bypassed with custom algorithms (Fig. 2).

## 4 Abstract layers design

When creating layouts for common CMOS nodes, typical electronic designers do not need to layout novel transistors or pCells. Indeed, an extensive library is usually provided by the CMOS manufacturer together with full description of the electrical characteristics, physical dimensions and schematic representation which are contained in the process design kit (PDK). If the designer requires more advanced cells, there is an option of accessing specialised libraries, such as those of ARM, which under a license agreement extend the designers capabilities. In rare cases however, when the solutions above remain insufficient, the designer is forced to create his own transistors, but this task is greatly simplified by the structure that manufacturers give to the design levels. For example, for defining the body of an nFET, it is sufficient to draw an n-well layer so that in a fab-internal post-processing step, a number of layers, such as stressors, halo and extension implants, block levels for some p-type implants, are automatically generated. In other words, the task of the low-level electronic designer is simplified by the structure of the design levels he can access. Moreover, mistakes are minimised, and proprietary process information can be conveniently hidden.

The framework described above however constitutes a challenge when designing photonic components in advanced microelectronics CMOS nodes (Fig. 3). As an example, while all

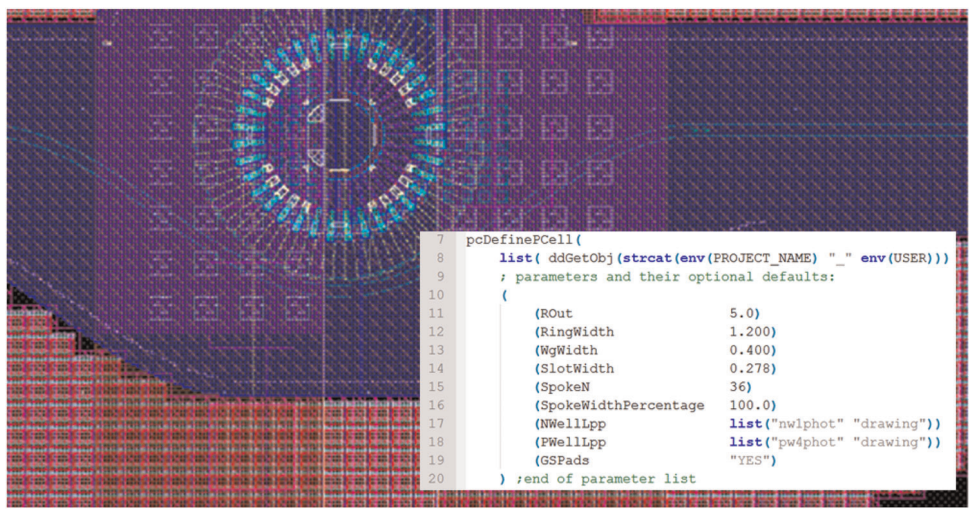

**Fig. 3** *Definition and representation of the pCell of a spoke-ring modulator*

On the top is recognisable the ring and the access waveguide; on the bottom is visible the dense metal fill pattern for avoiding density violations

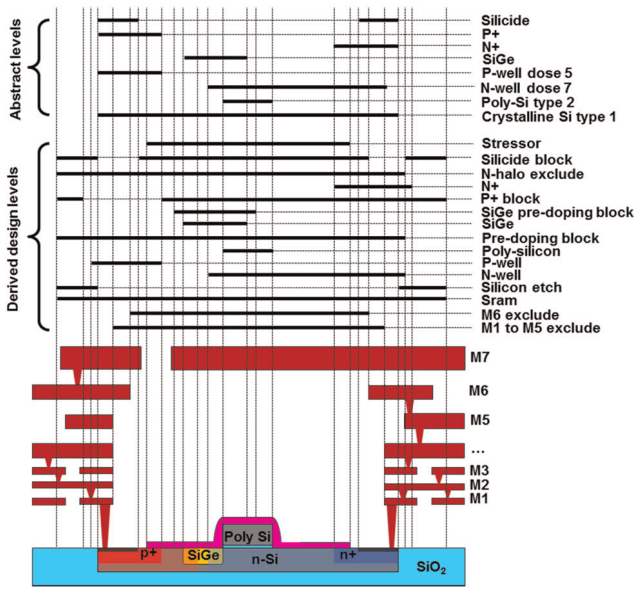

**Fig. 4** *Example of level generation*

Designer draws shapes on abstract levels which have a direct correspondence with the relevant parameters. Those layers are drawn irrespectively of DRs. For example, high-doping levels (N+ and P+) can be drawn with edges coincident with the silicon pattern. In a later data-preparation step, these layers are translated into counter-intuitive design levels and the DRs are enforced electronic devices require either n-type or p-type implants, photonics usually requires undoped silicon to avoid optical loss caused by free-carrier absorption. However, according to the process-flow of some electronics CMOS foundries, every time that a structure is not defined as n-well, it will automatically be doped p-type unless special design levels are accessed.

Although CMOS nodes have evolved over decades for improving and facilitating electronic design, photonic structures have not been taken into account during this development. As a consequence, design of photonics not only did not benefit from the methods developed, but has been made much more cumbersome by these very design flows. More specifically, the photonic designer would like to draw shapes on the individual masks used in the actual fabrication process: one would like to specify the location of crystalline silicon, stressors, ion-implants, poly-silicon and so forth. However the so-called 'mask levels' usually cannot be accessed directly. Instead, the designer is expected to draw shapes on 'design-levels' (such as the n-well described above) or 'utility-levels', Fig. 4. Utility levels are used for example to specify that certain locations are meant to be of a certain type, for example memory instead of logic cells – and this has consequences for the fabrication process. The photonics designer therefore will require drawing many layers, just for defining one particular feature, such as low optical-loss crystalline silicon.

As part of our design tool, we modified the original PDK of the CMOS foundry and added novel 'photonic' design levels which are technology-independent and have a universal meaning for photonics. Some of these layers represent for example undoped crystalline silicon, undoped poly-silicon, n-well implant alone and undoped silicon-germanium. Importantly, these layers have a tone (positive or negative) which more naturally corresponds to the intuition of the designer: in the case of doping, for example, one draws shapes corresponding to the locations which should be implanted instead than to the locations which should not receive the implants. The translation into final, technology-dependent design and utility layers happens during an automated data-preparation step, see Figs. 4 and 5.

## 5 CMOS DRs overview

The importance of satisfying certain geometrical constraints on mask designs for guaranteeing intended functionality and high yields was recognised since the early days of VLSI [29]. These constraints or DRs, define the allowed design patterns of individual or multiple design layers which are then converted into mask designs through a data preparation step which may involve, for example, optical-proximity correction. Common single-layer rules involve minimum/maximum area, space, length, notch or width or the allowed orientations of polygon edges. Examples of rules involving more layers are those defining the distance, extension or overlap between two or more layers.

The consequences of violating DRs can have different severities. Some, for example, are meant to guarantee the correct functionality of transistors and can therefore safely be waived when dealing with photonic devices. Other violations may affect the performance of the device locally, for example when silicide is not surrounded by a sufficient amount of highly-doped silicon which may lead otherwise to the formation of Schottky diodes. The most severe violations however may compromise the functionality of the entire wafer. Violations involving minimum-size rules are an example: if a layer feature is too small, the resulting too-narrow resist may detach from the wafer and reach distant locations, therefore compromising the functionality of other devices. Similarly, if two metal lines are drawn too close to each other they may short-circuit. Finally, also density requirements have to be met, for guaranteeing, for example, that chemo-mechanical polishing planarisation leads to acceptable thickness uniformity to prevent local dishing of the wafer. As an example, for achieving the

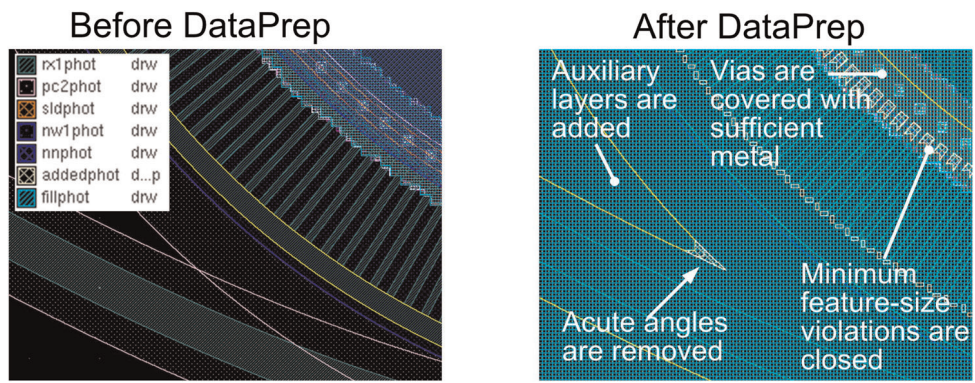

**Fig. 5** *Illustration of a layout before and after data-preparation*

On the left-hand side, is a spoked-ring with crystalline silicon (rx1phot) covered with poly-silicon (pc2phot), silicide (sldphot) and an n-well implant (nw1phot). On the right-hand side is the layout translated into technology-specific/auxiliary layers (a few tens), and design-rules violations are removed. In this example, acute angles are removed, minimum feature-size violations are closed and vias are covered with minimum metal. The access waveguide (bottom) and the ring may separately be DRC clean, but once put close to each other violations such as acute angles may appear: removal of DRC violations must occur after the placement of the individual devices minimum metal density requirements around the waveguides we introduce dense metal fill around their contour, Fig. 3. More subtle rules are those involving material properties. For example, if a silicon-germanium stressor is drawn too wide, it will not grow pseudomorphically on the silicon substrate, therefore leading to atomic dislocations and poor electrical (and optical) performance.

In standard CMOS nodes there are typically between 10 000 and 50 000 DRs. Although the majority of these rules have a meaning independent of whether the device is a transistor or an optical device, some rules apply to transistors only. Similarly, it is foreseeable that new rules will be required for photonic devices. For example, a silicon-germanium pocket can be utilised to make a pin-photodiode, and rules defining the overlap with n- and p-type dopants will be required. Other rules may for example prevent from placing absorbing materials, such as silicide, highly-doped silicon or metals, too close to optical waveguides.

In addition, connecting together two photonic components which are individually DRC-clean does not necessarily lead to a DRC-clean design. For example, these devices may require doping layers, exclude layers or fill layers which surround or extend beyond the silicon pattern, so that when the devices are connected together these layers may interfere with each other, see Figs. 5 and 6. As a consequence, the process of cleaning a design from design rule violations cannot happen on a device-level, but must happen on a chip-level. With the exception of small or simple designs, this is an effort which goes beyond the capability of individuals, and should be left to an automated algorithm.

## 6 Removal of DRC violations

As mentioned above, the removal of DRC violations is a process which cannot happen on a device-level, but must occur at the end of the design process, when all the devices are already in place. If the designer attempts to fix these violations manually and/or access directly the hundreds of design- and utility-levels without following any specific algorithm, he will probably go through a long series of DRC runs, iterating several manual fixes. This process is very time consuming and significantly limits the number of different cells one can design. For large scale integration, with, say, tens to hundreds of optical modulators and detectors all slightly tuned to create wavelength specific devices for a wavelength-multiplexed communication system, such manual approaches do not scale and make the task nearly impossible, and are certainly far from reliable.

Our tool is programmed to remove most DRC violations automatically during data-preparation, so that the designer is not required to have a deep understanding of all these rules, Fig. 5. The technology-dependent parameters are stored in a separate file, indicating all data specific to a particular process technology, for example, IBM 32 nm 13SOI, IBM 45 nm 12SOI or Texas Instruments 65 nm bulk CMOS. By simply replacing this file, a single design, made using the photonic levels above, will enable the generation of DRC-clean design levels across various technologies. Structures which appeared impossible to design in a zero-change CMOS approach, have now become feasible, as shown in Fig. 7, which illustrates a waveguide and ring resonator comprising subwavelength contacts on the sides.

## 7 Removal of minimum-size violations: the algorithm

In this section we will describe the entire workflow and focus on a single example of DRC cleaning: the removal of minimum-space violations.

The starting point is an arbitrarily large list of coordinates (even longer than 4000 points) which have been obtained, for example, by evaluating a mathematical function (such as a sine or a circle) or by integrating a differential equation as done with the spokes in Fig. 7. This list is then passed to a custom procedure which tiles the corresponding polygon into sub-polygons which have <4000 points (the absolute maximum tolerated by the format chosen by Cadence®) even after the Manhattanisation step. Next each list of coordinates is 'Manhattanised', meaning that the corresponding polygon is formed by orthogonal segments (on a Manhattan grid), Fig. 8. We typically choose a grid of 1 nm, which is fine enough for approximating any continuous boundary of a photonic component. The advantage of using orthogonal polygons is that the size operations (enlarging or diminishing the dimension of a polygon) are now well-defined. If $2d$ is the minimum space tolerated on a specific layer, this layer is first sized by $d$, Fig. 8c. This operation eventually closes opposite edges which are located closer than $2d$ from each other. When this shape is finally sized by $-d$, it does not return to the original form, but to a form where all opposite edges are separated by more than $2d$ –the DRC violation has been removed, Fig. 8d. After the data-preparation step, the standard DRC deck for electrical design provided by the CMOS foundry is executed for ensuring that all rules are correctly enforced. In addition, all shapes that have been added or removed automatically during the DR-cleaning step, are copied on special layers for further visual inspection. The majority of these shapes have moreover a dimension much smaller than the wavelength which does not impact the functionality of the photonic devices.

## 8 Waveguide ports, auto-routing and electrical pins

Photonic components are conveniently connected together by matching waveguides ports, a method implemented already by several photonic design packages. A waveguide port is defined as a list containing the port location, its orientation, its identifier (e.g. its name), its layer and its width. We implemented a SKILL procedure which allows connecting two ports together (therefore

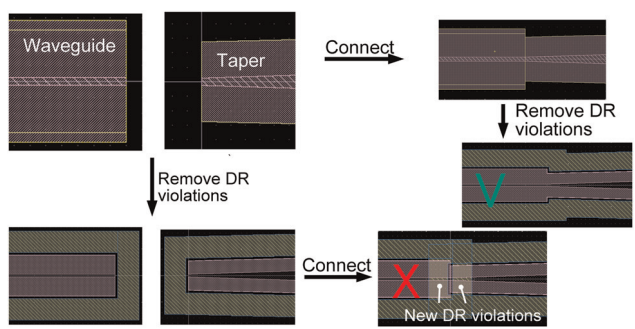

**Fig. 6** *Cleaning a design for a waveguide-to-taper transition from design rule violations must occur after connecting the components together*

Removing violations from connected devices (upper path) is not the same as connecting DRC-clean devices (lower path). In this example, DRC-clean optical waveguides must be surrounded by a fill layer that ensures that local material density rules are met. Overlap of such a layer with optical waveguides results in a DRC violation

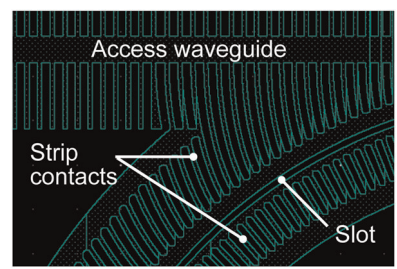

**Fig. 7** *Example of a DRC-clean slotted ring resonator with sub-wavelength silicon electrical contacts (strip)*

Strips are orthogonal to both the ring and the access waveguide, and are obtained by integrating a two-dimensional field

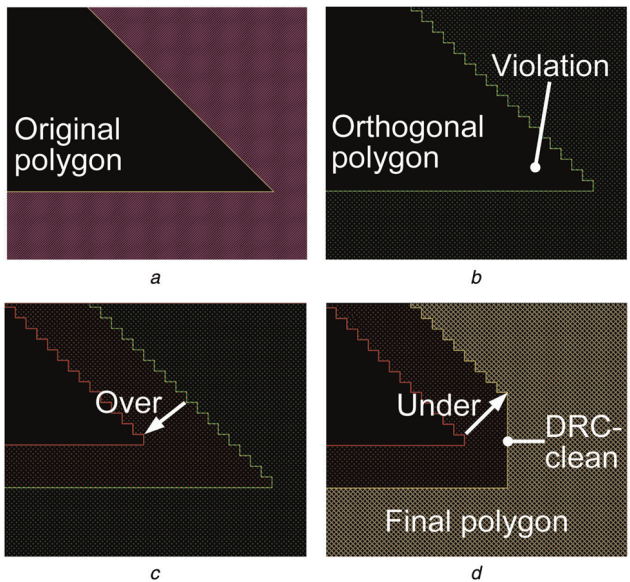

**Fig. 8** *Illustration of the algorithm for removing minimum-spacing design-rule violations*

*a* Original polygon
*b* Manhattanised polygon (orthogonal polygon)
*c* Size operation (positive amount) is made (over)
*d* Size operation by the same, but negative, amount is made (under). The minimum-space violation has disappeared. If the acute angle in (a) was formed by only three points and no Manhattanisation was used, the over-under operation would have left the shape unaltered. If the acute angle was formed by more points and still no Manhattanisation was used, the Over-Under operation would have led to unpredictable results

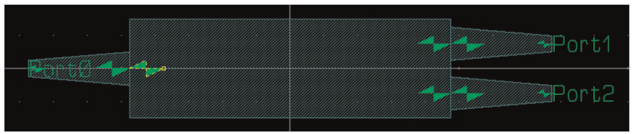

**Fig. 9** *Example of port functionality*

Multi-mode interference coupler is built staring from a rectangle and three tapers. Each of them has optical ports (bright green). The different parts are connected automatically by specifying port pairs performing the necessary translation and rotations of the objects) by just specifying the identity of the port pair, Fig. 9. We also implemented an auto-routing procedure which defines automatically a waveguide starting from the input and output ports. The code currently implements straight, sinusoidal and circular waveguides as most appropriate. The algorithm first defines a path connecting the ports (i.e. a one dimensional array of points) and then generates a waveguide from it using the 'path-to-waveguide' procedure mentioned above. Finally, standard SKILL-defined electrical pins are exploited on the optical devices as well, allowing Cadence® to perform electrical auto-routing as with normal transistors, Fig. 10. We therefore extended the functionality of Cadence® allowing performing both electrical and optical routing in a single environment.

## 9 Outlook

The PDA tool presented here suits especially the needs of early development phases, when the design is focused at improving the functionality of single devices without paying attention to system characteristics. At this stage, in fact, the designer sweeps the parameter space with many device variations, and this is best accomplished using a scripted language rather than editing through a graphical UI. The use of abstract layers and the following DRC-cleaning process allows the designer to focus on physical design rather than on technology details.

As the tools for PDA mature and once a consolidated device library is available, the designer will be able to generate complex systems and take advantage of optical netlists, photonic auto-routing, optical SPICE models and optical LVS checks which are currently being developed in the same framework. In the near future, a single system designer will be able to implement designs with many millions of electronic and photonic components seamlessly integrated during both design and manufacturing.

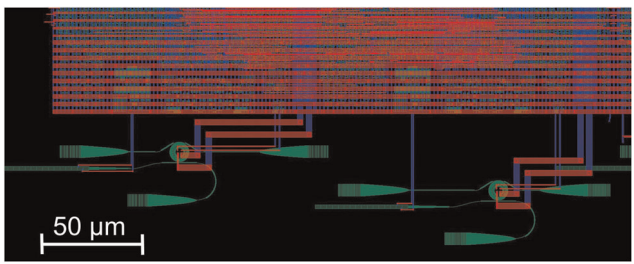

**Fig. 10** *Example of electrical and optical routing*

Electro-optic components, such as modulators and detectors, have electrical pins automatically routed to the microprocessor by Cadence®. Optical components are connected together using the photonic port and autorouting functionalities of our tool. The full processing time (for waveguide routing, layer generation and DR-cleaning) of the photonic components in this image is <80 seconds on a single logic CPU (out of the eight available) of a single intel Xeon X5550 processor at 2.7 GHz clock frequency

## 10 References

1  Assefa, S., Shank, S., Green, W., *et al.*: 'A 90 nm CMOS integrated nano-photonics technology for 25 Gbps WDM optical communications applications'. 2012 IEEE Int. Electron Devices Meeting (IEDM), 2012
2  Gunn, C.: 'Fully integrated VLSI CMOS and photonics 'CMOS Photonics'', 2007 IEEE Symp. on VLSI Technology, 2007, pp. 6–9
3  Young, I.A., Mohammed, E.M., Liao, J.T.S., *et al.*: 'Optical technology for energy efficient I/O in high performance computing', *IEEE Commun. Mag.*, 2010, **48**, pp. 184–191
4  Orcutt, J.S., Khilo, A., Holzwarth, C.W., *et al.*: 'Nanophotonic integration in state-of-the-art CMOS foundries', *Opt. Express*, 2011, **19**, pp. 2335–2346
5  Georgas, M., Moss, B.R., Sun, C., *et al.*: 'A monolithically-integrated optical transmitter and receiver in a zero-change 45 nm SOI process'. 2014 Symp. on VLSI Circuits Digest of Technical Papers, pp. 1–2
6  Shainline, J.M., Orcutt, J.S., Wade, M.T., *et al.*: 'Depletion-mode carrier-plasma optical modulator in zero-change advanced CMOS', *Opt. Lett.*, 2013, **38**, pp. 2657–2659
7  Imec: 'http://www2.imec.be/be_en/services-and-solutions/silicon-photonics.html'
8  Ihp: 'http://www.ihp-microelectronics.com/en/research/technology/si-photonics.html'
9  Leti: 'http://www-leti.cea.fr'
10  Tyndall: 'http://www.tyndall.ie'
11  Ime: 'http://www.a-star.edu.sg/ime'
12  Mosis: 'http://www.mosis.com'
13  Fiers, M., Lambert, E., Pathak, S., *et al.*: 'Improving the design cycle for nanophotonic components', *J. Comput. Sci.*, 2013, **4**, pp. 313–324
14  Bogaerts, W., Fiers, M., Dumon, P.: 'Design challenges in silicon photonics', *IEEE J. Sel. Top. Quantum Electron.*, 2014, **20**, pp. 1–8
15  Interconnect: 'http://www.lumerical.com/tcad-products/interconnect'
16  Ipkiss: 'http://www.ipkiss.org'
17  Luceda Photonics: 'http://www.lucedaphotonics.com'
18  PhoeniX: 'http://www.phoenixbv.com/index.php'
19  MentorGraphics: 'http://www.mentor.com'
20  Kallistos: 'http://www.photond.com/products/kallistos.htm'
21  Vpiphotonics: 'http://www.vpiphotonics.com/index.php'
22  WieWeb: 'http://www.wieweb.com'
23  Aspic: 'http://www.aspicdesign.com'
24  Condrat, C., Kalla, P., Blair, S.: 'Crossing-aware channel routing for integrated optics', *IEEE Trans. Comput.-Aided Des. Int. Circuits Syst.*, 2014, **33**, pp. 814–825
25  Pyxis: ' http://www.mentor.com/products/ic_nanometer_design/custom-ic-design'
26  Arriordaz, A., Bakker, A., Cao, R., *et al.*: 'Improvements in the silicon photonics design flow: collaboration and standardization'. 2014 IEEE Photonics Conf. (IPC), pp. 63–64
27  Orcutt, J.S.: 'Monolithic electronic-photonic integration in state-of-the-art CMOS processes'. PhD thesis, MIT 2012
28  Orcutt, J.S., Ram, R.J.: 'Photonic device layout within the foundry CMOS design environment', *IEEE Photon. Technol. Lett.*, 2010, **22**, pp. 544–546
29  Mead, C., Conway, L.: 'Introduction to VLSI systems' (Addison-Wesley, 1979)